\documentclass[twocolumn,showpacs,fleqn,nofootinbib]{revtex4}
\usepackage{amsmath}
\usepackage{graphicx}
\usepackage{float}
\usepackage{subfigure}
\usepackage{verbatim}
\usepackage{color}
\usepackage{comment}
\usepackage{hyperref}
\usepackage{epstopdf}
\usepackage{booktabs}
\usepackage[utf8]{inputenc}

\begin{document}
	
	\title{Investigating entanglement entropy at small-$x$  in DIS off protons and nuclei}
	\author{G.S. Ramos and M.V.T. Machado}
	
	\affiliation{High Energy Physics Phenomenology Group, GFPAE  IF-UFRGS \\
		Caixa Postal 15051, CEP 91501-970, Porto Alegre, RS, Brazil}

\begin{abstract}
In this work we analyze the entanglement entropy in deep inelastic scattering off protons and nuclei. It is computed based on the formalism where the partonic state at small-$x$ is maximally entangled  with proton being constituted by large number of microstates occurring with equal probabilities. We consider analytical expressions for the number of gluons, $N_{gluon}$,  obtained from gluon saturation models for the dipole-target amplitudes within the QCD color dipole picture. In particular, the nuclear entanglement entropy per nucleon is studied.  We also study the underlying uncertainties on these calculations and compare the results to similar investigations in literature.
\end{abstract}	

\maketitle
	
\section{Introduction}	

Recently, high energy physics community make strong efforts to use statistical physics concepts to describe the outcome of particle collisions \cite{Munier:2009pc,Iancu:2004es}. As an example, the central distribution of multiplicities of particle produced in such scatterings at high energies regime is related to the entropy produced by the collisions. In this context, one subject of study in recent years is the entanglement entropy \cite{Headrick:2019eth}, $S_{EE}$. It measures how far the particle system is from a pure quantum state. Specifically, the $S_{EE}$ quantifies the level of entanglement between different subsets of degrees of freedom in a quantum state. In an entangled system its quantum state can not be factored as a product of states of its local constituents. The confinement of quarks inside hadrons is a typical example of quantum entanglement as they are both correlated and not isolated objects. One way to probe the short distance structure inside hadrons is to consider hard scattering of deeply virtual photons off nucleons or nuclei. For large momentum transfer, small transverse distances of order $1/Q$ are probed by photons having virtualities $Q^2$. One place where this occurs is in deep inelastic scattering (DIS) of leptons off hadrons. The partons, i.e. quarks and gluons, constituting those hadrons are experimentally investigated for a long time and the kinematical range available by now for DIS off protons reaches $x\gtrsim 10^{-5}$ and $0.065 \lesssim Q^2\lesssim 10^5$ GeV$^2$ \cite{Diaconu:2008jj}. The Bjorken-$x$ variable is the longitudinal momentum fraction carried by these partons. Then, one question that arises is the  tension between a non-zero entropy resulting from different configurations of quasi-free incoherent partons and the zero von Neumann entropy for the probed hadron which is a pure state in its rest frame. One answer to this issue seems to be the quantum entanglement of partons \cite{Kharzeev:2017qzs}.  

The use of different theoretical techniques in quantum chromodynamics (QCD) in order to describe the entropy production and entanglement entropy of partons has been employed nowadays. For instance, by using the dominance of gluon fusion reaction in $t\bar{t}$ production at high energy colliders in Ref. \cite{Afik:2020onf} it was proposed the direct detection of entanglement by measuring the angular separation of their decay products (signature of spin-entanglement). Here, we summarize some key works in literature  related to these issues. In Ref. \cite{Peschanski:2012cw} the definition of dynamical entropy for dense QCD states of matter is proposed, which is   written as an overlap functional between the gluon distribution at different total rapidities and saturation radius, $R_s(x)=1/Q_s(x)$. The typical momentum scale in the saturated limit  is the saturation scale, $Q_s$. The formalism also has been extended to the initial preequilibrium state of a heavy ion collision.  The entanglement entropy between the two outgoing particles in an elastic scattering is presented in Ref. \cite{Peschanski:2016hgk} by using an $S$-matrix formalism taking into account partial wave expansion of the two-body states.  The identification of the physical origin of the divergence in the entropy expression appearing in \cite{Peschanski:2016hgk} and the its further regularization is done in Ref. \cite{Peschanski:2019yah}. The obtained finite $S_{EE}$ is then applied to proton-proton collisions at collider energies. On the other hand, the entropy of a jet is determined in Ref. \cite{Neill:2018uqw} by using the entropy of the hard reduced density matrix obtained from tracing over infrared states. The thermodynamical entropy associated with production of gluons is shown in Ref. \cite{Kutak:2011rb} taking into account unintegrated gluon distribution (UGD) based on saturation approach. One important conclusion is that the thermodynamical entropy behaves like multiplicity of produced gluons and there should exist an upper bound on entropy of gluons coming from the saturated sector of gluon UGD.
 In Ref. \cite{Hagiwara:2017uaz} the authors consider the entropy of quarks and gluons by using the Wehrl entropy, $S_W$, in QCD  which is the semiclassical counterpart of von Neumann entropy. They use the parton phase space  QCD Wigner and Husimi distributions, which are obtained from models that include gluon saturation effects.  The obtained Wehrl entropy is expressed in terms of the gauge invariant matrix element of the quark and gluon field operators. In asymptotic regime, $Y=\ln(1/x)\rightarrow \infty$ they found $S_W\propto Q_s^2(Y)\sim e^{\alpha Y}$, which agree in the same limit with the different definitions of entanglement entropy referred above \cite{Kutak:2011rb,Peschanski:2012cw}.

 Focusing particularly on entanglement entropy, it has been investigated  for soft gluons in the wave function of a fast hadron in Ref. \cite{Kovner:2015hga}. There, the entropy production in high energy collisions is also obtained in the context of color glass condensate (CGC) formalism for the hadron wavefunction. Along similar lines, in Ref. \cite{Armesto:2019mna} the authors define the CGC density matrix and present the evolution equations for this matrix (afterward, the effective density matrix was also analyzed in  detail in \cite{Li:2020bys}). These equations turn out to be similar to the Lindblad evolution. At large rapidities (high energies) the obtained $S_{EE}$ grows linearly with rapidity  both in the dilute and saturated regime driven by different rates. In Ref. \cite{Duan:2020jkz}, an entropy of ignorance, $S_I$, is introduced associated with the partial set of measurements on a quantum state. It is demonstrated that in the parton model the $S_I$ is equal to a Boltzmann entropy of a classical system of partons. Moreover, it was shown that the ignorance and entanglement entropies are similar at high momenta and distinct at the low ones\cite{Duan:2020jkz}. The main point rised there is that the lack of coherence and large entropy of partons must be due to the ability to measure only a restrict number of observables rather than to the entanglement of the observed partons with the degrees of freedom which are not observed as advocated in Ref. \cite{Kharzeev:2017qzs}.  
 
 Here, we focus on the work in Ref. \cite{Kharzeev:2017qzs}, where the von Neumann entropy of the parton system probed in DIS is derived within the nonlinear QCD evolution formalism. Then, this entropy is interpreted as the entanglement entropy between the spatial region resolved by $ep$ DIS and the rest of the proton. The authors shown that there is a simple connection between the gluon distribution, $xG(x,Q^2)$, and the $S_{EE}$ with all partonic microstates being equiprobable. In particular, at small-$x$, $S_{EE}(Y)=\ln[xG(Y,Q^2)]$, where in the limit of large $Y$ the entanglement entropy is maximal. In other words, the equipartitioning of microstates that maximizes $S_{EE}$ corresponds to the parton saturation. At asymptotic regime the entropy takes the form $S_{EE}\approx \frac{\alpha_sN_c}{\pi}\ln[r^2Q_s^2(Y)]\,Y$, with $r \sim 1/Q$ being the characteristic dipole size  in DIS. In Ref. \cite{Tu:2019ouv} an experimental test of these ideas was devised where the entropy reconstructed from the final state hadrons is compared to the entanglement entropy of the initial state partons. It is demonstrated that $S_h$ and $S_{EE}$ are in agreement at small-$x$ by using measured hadron multiplicity distributions at the Large Hadron Collider (LHC)

Motivated by those studies in this work we compute the entanglement entropy of partons within the nucleons and nuclei at high energies using analytical parametrizations for the gluon distribution function (PDF) based on parton saturation approach. In particular, the usual integrated gluon PDF, $xG(x,Q^2)$, is obtained from the unintegrated gluon distribution on the proton and nucleus using the correspondence between the color dipole picture and the $k_{\perp}$-factorization formalism in leading logarithmic approximation. We compare the results with the recent extractions of $S_{EE}$ from hadron multiplicities in DIS and proton-proton collisions at the LHC \cite{Tu:2019ouv}. In addition, we will cover kinematical ranges relevant for future lepton-hadron colliders like LHeC/FCC-eh \cite{Bruening:2013bga,Klein:2016uwv,Kuze:2018dqd} and eRHIC \cite{Accardi:2012qut}. Comparison with other approaches for parton entropy will be presented. We also determine the nuclear entanglement entropy per nucleon, $S_A$.  The paper is organized as follows. In next section, we start by briefly reviewing the calculation of the entropy of partonic density matrix which describes DIS within the parton model.  Given the proton wavefunction this matrix is obtained by reducing it with respect to the unobserved degrees of freedom and the $S_{EE}$ is identified with the von Neumann entropy. A comparison is done with other frameworks for computing parton entropy. In Sec. \ref{sec3} we present our main results and discuss the uncertainties and limitations of the approach. In last section  we summarize the main conclusions and perspectives.

\section{Theoretical formalism and comparison with other approaches}
\label{sec2}

\subsection{Parton entanglement entropy}
Here we follows the formalism presented in Ref. \cite{Kharzeev:2017qzs}, where the entanglement entropy is obtained in the framework of high energy QCD using both a simplified (1+1) dimensional model of nonlinear QCD evolution and a full calculation in (3+1) dimensional case described by the Balitsky-Kovchegov (BK) evolution equation. The main point is that the von Neumann (Shannon) entropy resulting from entanglement  between the two regions probed in DIS can be interpreted as the $S_{EE}$. The entropy is given by the Gibbs formula, $S_{EE}=-\sum_n p_n\ln(p_n)$, where $p_n$ is the probability of a state with $n$ partons.  Using a dipole representation, where a set of partons is represented by a set of color dipoles, the probability of microstates $p_n$  is identified with the probabilities to find $n$ color dipoles inside the proton at rapidity $Y$, $P_n(Y)$. In the toy model (1+1) dimensional, the latter quantity is obtained from the following relation of recurrence (dipole cascade equation),
\begin{eqnarray}
\frac{dP_n(Y)}{dY} = -n\,\alpha_hP_n(Y) + (n-1)\,\alpha_h P_{n-1}(Y),
\end{eqnarray}
where $\alpha_h$ is the BFKL intercept, $\alpha_h = 4\ln 2 \,\bar{\alpha}_s$ ($\bar{\alpha}_s=\alpha_s N_c/\pi$). This is quite similar to the Bateman equations for unstable nuclide decays, where the first term is due to the decay as $n\rightarrow (n+1)$ dipoles  and the second term corresponds to a growth rate dur to the splitting of dipoles, $(n-1) \rightarrow n$. In Ref.  \cite{Kharzeev:2017qzs}, the equation is solved by using the generating function technique, defining it as $Z(Y,u)=\sum_n P_n(u)u^n$. The initial conditions for dipole probabilities are  $P_1(0)=1$ plus $P_{n>1}(Y)=0$, with $\sum_nP_n(Y)=1$. These properties lead to the initial and boundary conditions to the generatrix function, $Z$. Assuming $Z(Y,u)=Z(u(Y))$, it can be shown that the dipole (parton) cascade equation includes nonlinear evolution in the form,
\begin{eqnarray}
\frac{\partial Z}{\partial  Y} & = & -\alpha_h(Z-Z^2),\\
Z(0,u)& = & u,\,\, Z(Y,1)=1,
\end{eqnarray} 
for rapidities near to those provided by the initial conditions. The differential equation for $Z(Y,u)$ at any rapidity is $\partial Z/\partial Y = -\alpha_h u(1-u)\partial Z/\partial u$.  By solving it  in this general case one obtains,
\begin{eqnarray}
Z(Y,u)=ue^{-\alpha_hY} \sum_{n=1}^{\infty} u^n (1-e^{-\alpha_hY})^n.
\end{eqnarray}

Rewriting the solution in terms of $P_n$, finally one obtains,
\begin{eqnarray}
P_n(Y) = e^{\alpha_h \,Y}\left(1-  e^{\alpha_h \,Y}\right)^{n-1}.
\end{eqnarray}

By doing the identification $p_n=P_n(Y)$ and using the Gibbs formula, the von Neumann entanglement entropy as a function of $Y$ reads as:
\begin{eqnarray}
\label{1p1}
 S_{EE}(Y)=e^{\alpha_h Y}(\alpha_hY)+\left(1- e^{\alpha_h Y} \right)\ln \left(e^{\alpha_h Y}-1\right),
\end{eqnarray}
which presents the following limit, $S_{EE}(\alpha_hY \gg 1) \sim \alpha_hY$.

 By defining the gluon distribution, $xG(x,Q^2)$, as the average number of partons, $\langle n\rangle$, probed with resolution $Q^2$ at a given value of $x$, one obtains,
 \begin{eqnarray}
 \langle n\rangle=\sum_n nP_n(Y) = u\left. \frac{\partial Z(Y,u)}{du}\right|_{u=1}=e^{\alpha_hY}.
 \end{eqnarray}
 
 Comparing the average number of gluon $\langle n\rangle=x^{-\alpha_h}$ and the entropy expression in Eq. (\ref{1p1}) the following relation is obtained at the limit $\alpha_hY\gg 1$,
\begin{eqnarray}
 S_{EE}=\ln \left[xG(x,Q^2)  \right],
 \label{xgsee}
\end{eqnarray}
which is a key result presented in \cite{Kharzeev:2017qzs}. The limit  $\alpha_hY\gg 1$ is satisfied by values of Bjorken-$x$ less than $\sim 10^{-3}$.  The von Neumann entropy was obtained from the reduced density matrix $\hat{\rho}_A= \mathrm{Tr}_B \hat{\rho}_{AB}$ (partial trace), where the proton probed in DIS is considered as a bi-partite system ($A$ is the region of space probed in the hard process and B is the one complementary to A, i.e. the rest of proton). The wavefuntion of this bi-partite system is constructed based on the orthonormal set of states, $|\psi_n^A\rangle$ and $|\psi_n^B\rangle$ by using Schmidit decomposition, $|\psi_{AB}\rangle= \sum_nc_n|\psi_n^A\rangle |\psi_n^B\rangle$. The authors of Ref. \cite{Kharzeev:2017qzs} assume that the full set of states is defined by the Fock states with distinct numbers $n$ of partons. Therefore, $\hat{\rho}_A = \sum_nc_n^2|\psi_n^A\rangle \langle \psi_n^A|$, where $c_n^2\equiv p_n$ is identified with the probability of a state of $n$ partons. 

The calculation for a full (3+1) dimensional QCD is more involved. The starting point is writing down the parton cascade equation whose solution gives the probability to have $n$-dipoles, $P_n(Y;\{r_i\})$ (with the notation, $\{r_i\}=r_1,r_2, \ldots, r_i,\ldots, r_n$), at rapidity $Y-y$ and transverse size $r_i$. The cascade equation conducts to the BK evolution equation for dipole amplitude and takes the form,
\begin{eqnarray}
\label{cascfull}
\frac{dP_n(Y;\{r_i\})}{dY} & = & -\sum_{i=1}^{n}\bar{\alpha}_s \omega (r_i) P_n(Y;\{r_i\})  \\
& +& \sum_{i=1}^{n-1} K(r_i,r_n|\vec{r}_i+\vec{r}_n) P_{n-1}(Y;\{(\vec{r}_i+\vec{r}_n)\}), \nonumber
\end{eqnarray}
where $P_n$ obeys a sum rule and similar initial condition as the $(1+1)$ case,
\begin{eqnarray}
\sum_{n=1}^{\infty} \int \prod_{i=1}^{n}d^2\vec{r}_i P_n(Y;\{r_i\}) &=&1,\\
P_{n>1}(0;\{r_i\}) & = &0,\\
P_1(Y;r_1) &=& \delta^{(2)}(\vec{r}-\vec{r}_1)e^{-\omega (r_1)\bar{\alpha}_s Y}.\nonumber
\end{eqnarray}

The surviving probability of one dipole is given by $\bar{\alpha}_s \omega (r_i)=\bar{\alpha}_s \ln(r_i/\mu^2)$, with $\mu^2$ being an infrared cutoff.  The probability of a dipole having size $|\vec{r}_i+\vec{r}_n|$  to decay into two with the transverse sizes $r_i$ and $r_n$ is given by,
\begin{eqnarray}
K(r_i,r_n|\vec{r}_i+\vec{r}_n) = \frac{\bar{\alpha}_s }{2\pi}\frac{(\vec{r}_i+\vec{r}_n)^2}{r_i^2r_n^2}.
\end{eqnarray}

In Ref. \cite{Kharzeev:2017qzs} the parton cascade equation is solved by using the Mellin transform technique (with $\omega$ being the conjugate variable to $Y$), which  produces the following,
\begin{eqnarray}
P_n(Y;\{r_i \})& = & \int_{\epsilon-i\infty}^{\epsilon+i\infty} \frac{d\omega}{2\pi}e^{\omega \bar{\alpha}_s Y} {\cal{P}}_n(Y;\{r_i\}), \\
{\cal{P}}_n(\omega, \{r_i\}) &=& 2\pi r^2 \delta^{(2)}(\vec{r}-\vec{r}_1)\left( \frac{1}{2\pi} \right)^n\prod_{i=1}^{n}\frac{\Omega_n(\omega;\{r_i\})}{r_i^2},\nonumber \\
\omega \Omega_n(\omega,\{\omega_i\}) & \equiv & -\left( \sum_{i=1}^{n}\omega_i \right)\Omega_n(\omega,\{\omega_i\}) \nonumber\\
&+& \sum_{j=1}^{n-1}  \Omega_{n-1}(\omega,\{\omega_i,\omega_{jn}\}), \nonumber
\end{eqnarray}
with the notation $\omega_i=\omega (\vec{r}_i)$ and $\omega_{ij}=\omega (\vec{r}_i+\vec{r}_j)$. 

The general solution, given in terms of a recurrence formula for the  $\Omega_n$ function takes the form,
\begin{eqnarray}
\Omega_n(\omega,\{\omega_i\})=  (n-1) \frac{\Omega_{n-1}(\omega,\{\omega_i,\omega_{n-1,n}\})}{\omega + \sum_{j=1}^{n}\omega_j},
\end{eqnarray}
which is solved explicitly in \cite{Kharzeev:2017qzs} for two special cases: (a) for $r_i,r_n\gg |\vec{r}_i+\vec{r}_n |$  (corresponding to the perturbative QCD double logarithm approximation, DLA) and (b)   $|\vec{r}_i+\vec{r}_n |\rightarrow r_i$ while $r_n\ll r_i$ (corresponds to parton cascade evaluated in the saturation region). In the latter case, the solution is given by $\Omega_n(\omega,\{\omega_i\})=(n-1)! \prod_{j=1}^{n}(\omega + \sum_{\ell=1}^{n} z_{\ell})^{-1}$ where $z_{\ell}\equiv \omega_i=\ln(r_i^2Q_s^2)$. Accordingly, the solution in this case is given by,
\begin{eqnarray}
P_n(Y;\{r_i \})& = & 2\pi r^2 \delta^{(2)}(\vec{r}-\vec{r}_1)\left( \frac{1}{2\pi} \right)^n\prod_{i=1}^{n} \frac{1}{r_i^2}\nonumber \\
&\times & \int_{\epsilon-i\infty}^{\epsilon+i\infty} \frac{d\omega}{2\pi}e^{\omega \bar{\alpha}_s Y} \Omega_n(\omega,\{z_i\}) ,\\
&=& 2\pi r^2 \delta^{(2)}(\vec{r}-\vec{r}_1)\left( \frac{\bar{\alpha}_s Y}{2\pi} \right)^n\prod_{i=1}^{n}\frac{1}{r_i^2} \nonumber \\
&\times & e^{-\bar{\alpha}_s z_1Y}\prod_{i=2}^n \Phi (t_i),
\end{eqnarray}
where $t_i=\bar{\alpha}_sY\sum_{\ell =i}^{n}z_{\ell}$ with $\Phi (t_i) = (1-e^{t_i})/t_i$. Using this solution, the following relation can be evaluated analytically,
\begin{eqnarray}
\int\prod_{i=1}^{n} d^2\vec{r}_iP_n(Y; \{r_i\}) &=&   \int_{\epsilon-i\infty}^{\epsilon+i\infty} \frac{d\omega}{2\pi}e^{\omega \bar{\alpha}_s Y}\int \prod_{i=1}^{n}dz_i \Omega_n,\nonumber\\
&=&  \frac{1}{n !}\Xi^n(\bar{\alpha}_sz_1 Y)e^{-\bar{\alpha}_sz_1 Y},
\end{eqnarray}  
where the auxiliary function $\Xi$ in equation above takes the form,
 \begin{eqnarray}
  \Xi (t_n)  =  \int_0^{t_n}\Phi(t)dt=\gamma_E+\Gamma (0,t_n)+\ln (t_n), 
 \end{eqnarray}
with $\gamma_E$ being the Euler-Mascheroni constant and $\Gamma (0,t)$ is the incomplete gamma function. 

In the limit of large rapidity $Y$, with $\bar{\alpha}_s z_1 Y\gg1$,  the entanglement entropy in (3+1) QCD is evaluated as (see Ref. \cite{Kharzeev:2017qzs} for details),
\begin{eqnarray} 
\label{1p3}
S_{EE}&=&-\sum_{n=1}^{\infty}\prod_{i=1}^{n}\int d^2r_iP_n(Y;{r_i})\,\ln \left[P_n(Y;{r_i}) \right],\nonumber \\
&=& \Delta_s Y-e^{-\Delta_s Y}\int_0^{\Delta_s Y}t_n\Phi (t_n)e^{\Xi (t_n)}dt_n, 
\end{eqnarray}
 where one defines $\Delta_s =\bar{\alpha}_s\ln(r^2Q_s^2)$  with $r$ being the typical dipole size in DIS and $Q_s(x)$ is the saturation scale.The second term is subleading for any rapidity and in the limit of large $Y$ (very small-$x$) the entropy  has the asymptotic form $S_{EE}\approx \Delta_s Y$.  The latter has the same behavior that the $(1+1)$ calculation by the replacement $\alpha_h\rightarrow \Delta_s$.
 
Now, we introduce our contribution to the theme.  Here, we will take into account an analytical expression for the gluon PDF \cite{GolecBiernat:1999qd}, which is valid for $Q^2\leq 50$ GeV$^2$ and allow us to obtain the number of gluons down to very small virtualities, $Q^2\ll 1$ GeV$^2$. This is an advantage compared to the usual PDFs extracted from fitting initial conditions at $Q^2=Q_0^2\approx 2$ GeV$^2$ and further DGLAP evolution. Another advantage is that it is an explicit function of the saturation scale, $Q_s(x)$. Starting from the GBW saturation model \cite{GolecBiernat:1999qd}  which nicely describes all data on $F_2$, $F_L$, exclusive vector meson production and diffractive structure function in the small-$x$ regime one obtains the unintegrated gluon distribution (UGD), $\alpha_s{\cal F}(x,k_{\perp}) = N_0\,(k_{\perp}^2/Q_s^2)\exp(-k_{\perp}^2/Q_s^2)$, with $N_0 = 3\sigma_0/4\pi^2$. The usual integrated gluon PDF can be calculated from the UGD, 
\begin{eqnarray}
 xG(x,Q^2) & = & \int_0^{Q^2} dk_{\perp}^2{\cal F}(x,k_{\perp}), \nonumber \\
 & = & \frac{3\sigma_0}{4\pi^2\alpha_s}Q_{s}^2\left[1-\left(1+\frac{Q^2}{Q_{s}^2} \right)e^{-\frac{Q^2}{Q_{s}^2}} \right],
\end{eqnarray}
where $Q_s(x)=(x_0/x)^{\lambda/2}$ gives the  transition between the dilute and saturated gluon system. In numerical calculations in next section, we use the updated values for the model parameters (fit result including charm):  $\sigma_0=27.32$  mb, $\lambda = 0.248$  and $x_0=4.2\times 10^{-5}$   \cite{Golec-Biernat:2017lfv}. The presence of the nucleon saturation scale will be useful when investigating the entropy for DIS off nuclei as discussed in subsection \ref{sec3b}.

Before doing our phenomenological analyses in next section, we would like to contrast the entanglement entropy proposed in  \cite{Kharzeev:2017qzs} to other formalisms for evaluation of parton entropy. In next subsection, we discuss the main results coming from the entanglement entropy (von Neumann) computed in the Color Glass Condensate formalism  and from the semiclassical Wehrl entropy for gluons. 

\subsection{Comparison to other frameworks}
\begin{figure}[t]
		\includegraphics[scale=0.35]{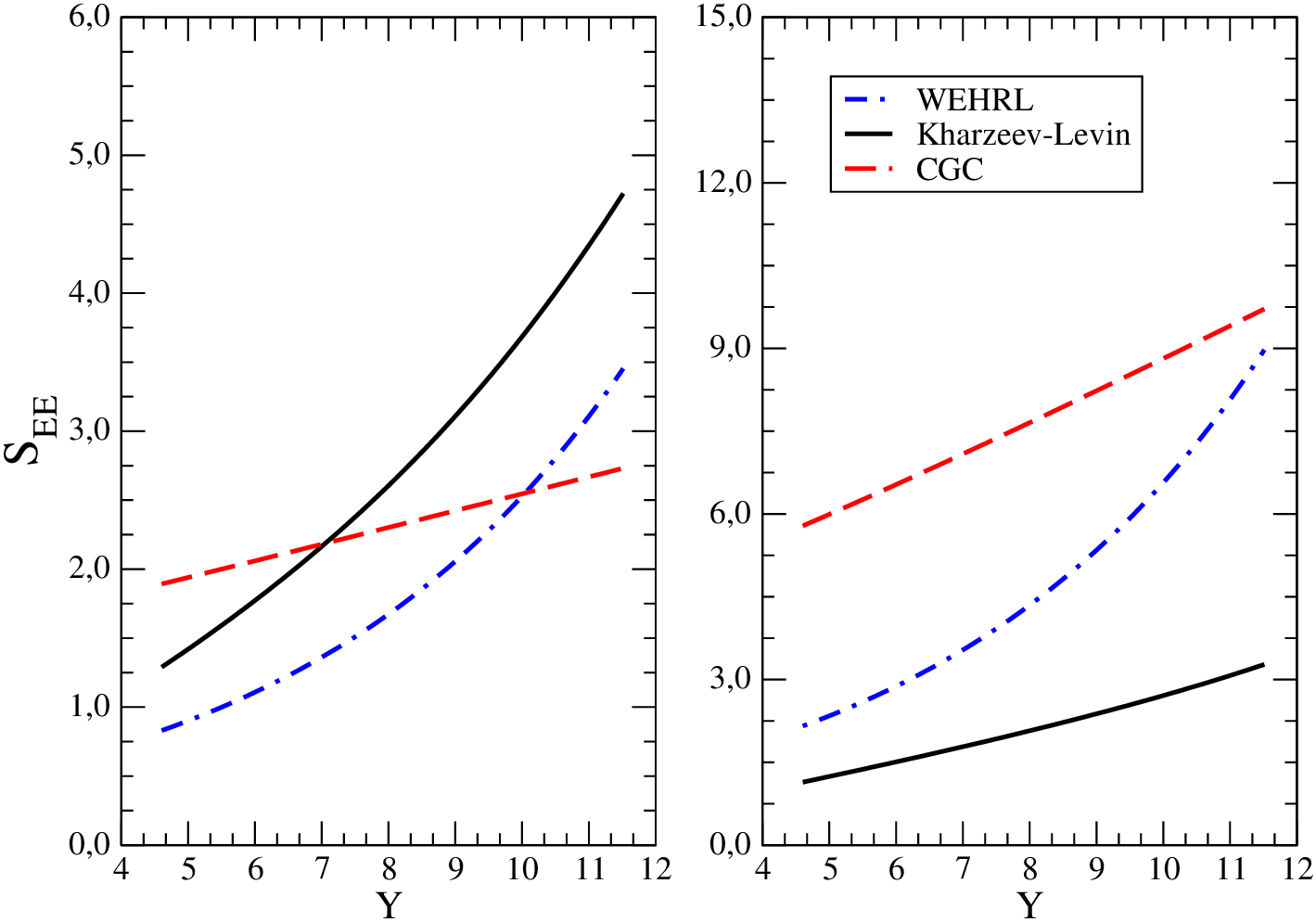} 
	\caption{The comparison of different approaches for the parton (gluon) entropy at small-$x$. The entropy is plotted as a function of $x$ for virtualities $Q^2 =2$ and $10$ GeV$^2$ in DIS off protons. Results are shown for entanglement entropy by Kharzeev-Levin, the entanglement entropy in the CGC formalism and the Werhl entropy for gluons.}
	\label{fig:0}
\end{figure}

In this subsection, we compare the entanglement entropy discussed above to other formalisms for the entropy of parton states at high energy limit. We start with the entanglement entropy evaluated in the context of color glass condensate approach (CGC). It is obtained by taking into account  soft gluons (i.e., gluon field modes with small longitudinal momenta) in the wavefunction of a fast moving hadron. In Ref. \cite{Kovner:2015hga}, the reduced density matrix for these soft modes is computed in the McLerran-Venugopalan (MV) model, $S_{EE} = -\mathrm{Tr} [\hat{\rho}_{\mathrm{MV}}\ln\hat{\rho}_{\mathrm{MV}}]$. The reduced $\hat{\rho}$ is written in terms of the matrix elements of a matrix $M_{ij}^{ab}\propto g^2/4\pi^2\int dudv \mu^2(u,v)(x-u)_i(y-u)_j\delta^{ab}$ (see \cite{Kovner:2015hga} for details), with $g$ being the strong coupling. By using translational invariance it is a function of soft gluon transverse momentum, $k$, with  $  M_{ij}^{ab}(p)=g^2\mu^2(p) (p_ip_j/p^2)\delta^{ ab}$. In MV model, the quantity $\mu^2$ is independent of $k$ and related to the  gluon saturation scale. Authors obtained parametric solutions for $S_{EE}$ from both large (UV modes) and small (IR modes) transverse momenta by expanding $M$ accordingly. Namely,
\begin{eqnarray}
\label{cgc1}
S_{EE}^{\mathrm{UV}}&\propto  & \frac{1}{2}S_{\perp}(N_c^2-1)\int \frac{d^2k}{(2\pi)^2} \frac{g^2\mu^2}{k^2}\ln \left(\frac{k^2}{g^2\mu^2} \right),  \\
\label{cgc2}
S_{EE}^{\mathrm{IR}}&\propto &  \frac{1}{2}S_{\perp}(N_c^2-1)\int \frac{d^2k}{(2\pi)^2} \ln \left( \frac{g^2\mu^2}{k^2}  \right),
\end{eqnarray}    
where $S_{\perp}$ is the total area of nucleon/nucleus projectile and color factor $(N_c^2-1)=2N_c C_F$ appears as the density matrix is a product of density matrices over the color index. The gluon saturation scale is identified as $\bar{Q}_s^2 = g^4\mu^2$  and the large momentum integration is logarithmicaly divergent and it is regulated by a UV cutoff, $\Lambda$. The leading contributions for Eqs. (\ref{cgc1}-\ref{cgc2}) are found to have the form,
\begin{eqnarray}
\label{cgc3}
S_{EE}^{\mathrm{UV}}&\approx  & \frac{1}{2}S_{\perp}(N_c^2-1) \frac{\tilde{Q}_s^2}{2\pi g^2}\left[\ln \left(\frac{g^2\Lambda^2}{\tilde{Q}_s^2} \right) + \ln^2 \left(\frac{g^2\Lambda^2}{\tilde{Q}_s^2} \right)  \right],  \nonumber\\
\label{cgc4}
S_{EE}^{\mathrm{IR}}&\approx &  \frac{1}{2}S_{\perp}(N_c^2-1)\frac{3\tilde{Q}_s^2}{4\pi g^2}.
\end{eqnarray}

The calculation above performed  in field basis has been also done in the number representation basis in Ref. \cite{Duan:2020jkz}. They are shown to be coincident. The full expression for the von Neumann (entanglement)   entropy in number basis is given by,
\begin{eqnarray}
S_{EE} & \approx & \frac{1}{2}S_{\perp}C_F \int_0^{\Lambda^2}\!\!\frac{d^2k}{(2\pi)^2}\left[ \ln \left(\frac{g^2\mu^2}{k^2}\right) + \sqrt{1+ 4\frac{g^2\mu^2}{k^2}} \right. \nonumber \\
& \times & \left. \ln \left(1+\frac{k^2}{2g^2\mu^2} +\frac{k^2}{2g^2\mu^2}\sqrt{1+ 4\frac{g^2\mu^2}{k^2}}\right) \right].
\end{eqnarray}

In \cite{Kovner:2015hga,Duan:2020jkz} only qualitative parametric expressions are analyzed and no numerical calculations are presented for  $S_{EE}$. Here, we intend to do some phenomenology. The calculation above is performed at fixed rapidity and the UV cutoff is not specified\footnote{The evolution of entanglement entropy as a function of on the hadron rapidity in weak coupling case can be computed using a convolution of evolution equation kernels (BFKL, BK) with the gluon UGD.}. For phenomenological purposes we consider that the saturation scale can evolve with rapidity, $Y=\ln (1/x)$, following the GBW ansatz, $\tilde{Q}_s^2(x)=(9/4)(x_0/x)^{\lambda}$. Moreover, we will identify the UV regulator by the photon virtuality in DIS, with the arbitrary choice $Q^2 = g^2\Lambda^2$. We have computed analytically the  integration above, which takes the form,
\begin{eqnarray}
\label{CGCSEE}
S_{EE}^{\mathrm{CGC}}&=&\frac{1}{2}S_{\perp}\frac{C_F}{4\pi}\tilde{Q}_s^2\left[ \tau\ln \left(\tau^{-1}\right) +\tau\sqrt{1+4\tau^{-1}} \right.\nonumber \\
&\times & \left.  \ln \left( \frac{\sqrt{1+4\tau^{-1}}+1}{\sqrt{1+4\tau^{-1}}-1} \right)+ \ln^2\left( \frac{\sqrt{1+4\tau^{-1}}+1}{\sqrt{1+4\tau^{-1}}-1} \right)\right],\nonumber \\
\end{eqnarray}
where $\tau=Q^2/\tilde{Q}_s^2$. The parametric behaviors of Eqs. (\ref{cgc4}) are properly obtained, since for $\tau=1$ ($Q^2=\tilde{Q}_s^2$) then $S_{EE}\sim S_{\perp}\tilde{Q}_s^2$. On the other hand, for large $\tau$ ($Q^2\gg \tilde{Q}_s^2$), $\sqrt{1+4\tau^{-1}}\approx 1+ (2\tau^{-1})$ and thus we easily get $S_{EE}\sim S_{\perp}\tilde{Q}_s^2[2\ln(\tau)+\ln^2(\tau)]$. For numerical calculations we will use the parameter $S_{\perp}=\pi R_p^2=\sigma_0/2$ and GBW parameters for calculating the saturation scale as a function of rapidity.

Another formalism we will address is the Wehrl entropy in QCD \cite{Hagiwara:2017uaz}, which is the semiclassical analogue of the von Neumann entropy. It is obtained in terms of phase space distributions. In our context here,  one considers the multidimensional  QCD Wigner phase space distributions for gluons at small-$x$.  One advantage is that the entropy for quarks can be also computed in the same formalism, which is more general and model independent that the approaches considered before. The QCD Wigner distribution, $W$, is a generalization of the usual collinear parton distribution functions. Namely, it depends on parton transverse momentum, $\vec{k}$,  impact parameter, $\vec{b}$, and longitudinal parton momentum fraction, $x$. By integrating the Wigner distribution on the complete phase space, the usual PDFs are recovered.  If the Wigner distribution is positive definite (and not strongly oscillating) for the parton considered, then the Wehrl entropy can be defined,
\begin{eqnarray}
S_W &= &  -\int d^2bd^2k\, xW_{q,g}(x,k,b)\ln \left[ xW_{q,g}(x,k,b) \right], \nonumber\\
xf_{q,g}(x)& = & \int d^2bd^2k \,xW_{q,g}(x,k,b),
\label{WehrlS}
\end{eqnarray}
where $xf_q(x) = xq(x)$ and $xf_g(x)=xg(x)$ are the collinear distributions for quarks and gluons, respectively.

For our purpose, we will consider the Weiszacker-Williams (WW) gluon Wigner distribution\footnote{The dipole Wigner distribution for gluons has been derived in Ref. \cite{Hatta:2016dxp}.}, which can be computed in a quasiclassical approximation \cite{Kovchegov:1998bi,Dominguez:2011wm}. It is written in terms of the forward  $S$-matrix of a QCD color dipole of transverse size $\vec{r}$, transverse momentum $\vec{k}$ at impact parameter $\vec{b}$ in the adjoint representation, ${\cal{S}}_A$,
\begin{eqnarray}
xW_g(x,k,b) = \frac{C_F}{2\pi^4 \alpha_s}\int d^2\vec{r} \,\frac{e^{i\vec{r}\cdot \vec{k}}}{r^2}\left( 1-{\cal{S}}_A(x,\vec{r},\vec{b}) \right). \nonumber\\
\end{eqnarray}

The WW Wigner distribution can be analytically evaluated in the case of a Gaussian form for S-matrix, ${\cal{S}}_A(x,r,b)=\exp [-\vec{r}^2\tilde{Q}_s^2(x,b)/4]$, where $\tilde{Q}_s^2(x,b) =(N_c/C_F)Q_s^2(x,b)$ is the impact parameter dependent  gluon saturation scale. Specifically, for the Gaussian $S$-matrix one obtains,
\begin{eqnarray}
xW_g(x,k,b) = \frac{C_F}{2\pi^3\alpha_s}\Gamma\left( 0,\frac{k^2}{\tilde{Q}_s^2(x,b)}\right),
\label{WW}
\end{eqnarray}
which is positive definite with $\Gamma$ being the incomplete gamma function. Putting expression of  Eq. (\ref{WW}) in the definition of Wehrl entropy associated to the Wigner distribution, Eq. (\ref{WehrlS}), and disregarding overall prefactor of $xW$ in the logarithm, the entropy  $S_W$ can be obtained. We see that the integrand is a function of the ratio $\tau_k=k^2/\tilde{Q}_s^2$ and this fact helps the integration over transverse momentum. Here, in order to introduce a dependence on the resolution scale we replace the upper limit on $k$-integration by $Q^2$ instead of infinity. After change of variables, $k\rightarrow \tau_k$, the entropy reads as, 
\begin{eqnarray}
S_W & =&  -\frac{C_F}{2\pi \alpha_s}\int_{0}^{\infty}db^2\, F(\tau)\tilde{Q}_s^2(x,b),\\
F(\tau)&=& \int_0^{\tau}d\tau_k \,\Gamma(0,\tau_k)\ln \Gamma \left( 0, \tau_k\right),
\end{eqnarray}
with $\tau=Q^2/\bar{Q}_s^2(x,b)$. Putting $Q^2$ (and for consequence, $\tau$) to infinity, the function $F$ is just a number, $F(\tau\rightarrow \infty)\approx -0.248$. Notice that for finite $Q^2$, $F$ is a function of both $x$ and impact-parameter. For numerical calculations, we will use the impact-parameter(quark)  saturation scale from the b-CGC model \cite{Rezaeian:2013tka}, where $Q_s^2(x,b)=(x_0/x)^{\lambda}\exp [ -b^2/2\gamma_s B_{\mathrm{CGC}} ]$. The parameters are fitted to small-$x$ DIS data, with $x_0=0.00105$, $\lambda = 0.2063$, $\gamma_s=0.6599$ and $B_{\mathrm{CGC}}=5.5$ GeV$^{-2}$ \cite{Rezaeian:2013tka}. For simplicity, to avoid to compute numerically the impact parameter integration we take into account that the saturation scale has a maximum at $b=0$, with $\tilde{Q}_{s,max}^2(x)=\tilde{Q}_{s}^2(x,b=0)=(N_c/C_F)(x_0/x)^{\lambda}$. Moreover, in the small-$x$ region the typical saturation scale is of order 1 GeV  or so (using the b-CGC for the $x_0$  parameter, the  quark saturation scale  is of order unity around $x=10^{-3}$). Therefore, in our evaluations of $S_W$ we will use $\tau = Q^2/\langle \tilde{Q}_s^2 \rangle$ with $\langle \tilde{Q}_s^2 \rangle = 1 $ GeV$^2$. This gives $F\approx -0.095377$ for $Q^2=2$ GeV$^2$ and  $F\approx -0.247802$ for $Q^2=10$ GeV$^2$. For any $Q^2$, after integration on impact parameter, one has for the b-CGC model for the impact parameter saturation scale,
\begin{eqnarray}
\label{SWap}
S_W  \approx  - \frac{2F\gamma_s B_{\mathrm{CGC}}N_c}{2\pi \alpha_s}\,Q_s^2(x)= -\frac{2FN_c S_{\perp}}{6\pi^2\alpha_s}\,Q_s^2(x),
\end{eqnarray}
where the quantity $B_G=\gamma_s B_{\mathrm{CGC}}$ is related to the electromagnetic proton radius $R_p^2= 3B_G$ (with $S_{\perp}=\pi R_p^2$). Thus, the parametric behavior of the Werhl entropy obtained from the WW Wigner gluon distribution is $S_W\propto S_{\perp}Q_s^2(x)$.

Comparing the distinct approaches for entropy for gluons at small-$x$ we see that both CCC entanglement entropy, Eq. (\ref{CGCSEE}),  and the Wehrl entropy, Eq. (\ref{SWap}), are proportional to the transverse area of the target. This is an intrinsic property of any extensive observable as the entropy and the corresponding consequence for nuclear targets  will be adressed in next section. Such a property is not present in the parametric expression for the entanglement entropy proposed in \cite{Kharzeev:2017qzs} (KL), Eq. (\ref{1p3}). In Fig. \ref{fig:0} a comparison is done between the different evaluations for the gluon entropy. It  is plotted as a function of $x$ for virtualities $Q^2 =2$ (left panel) and $10$ GeV$^2$ (right panel). We set $\alpha_s=0.25$ in the calculations. Results are shown for entanglement entropy by Kharzeev-Levin (solid lines), the entanglement entropy in the CGC formalism (long dashed lines) and the Werhl entropy (dot-dashed lines). The parametric expression of KL model behaves like $S\sim Y^2$, with a logarithmic suppression in $1/Q^2$ as seen in figure. The choice $r^2=(4/Q^2)$ for the average dipole size was used and for the product inside logarithm one has $Q_s^2r^2=(4Q_s^2/Q^2)+e$ ( the second term is to prevent negative values of the argument when $Q_s^2\ll Q^2$).  On the other hand, The Wehrl entropy behaves like $S_W\sim e^{\lambda Y}$ and grows with $Q^2$ in our simplification of the $k$-integration, which is enough for the phenomenological purpose presented here. Now, the CGC entropy behaves as $S_{CGC}\sim e^Y[\ln^2 Q^2 - (2\lambda)Y]$, which explains the mild growth on $Y$ in figures.

\section{Results and discussions}
\label{sec3}
\subsection{Entanglement entropy for hadrons}
\label{sec3a}

\begin{figure}[t]
		\includegraphics[scale=0.35]{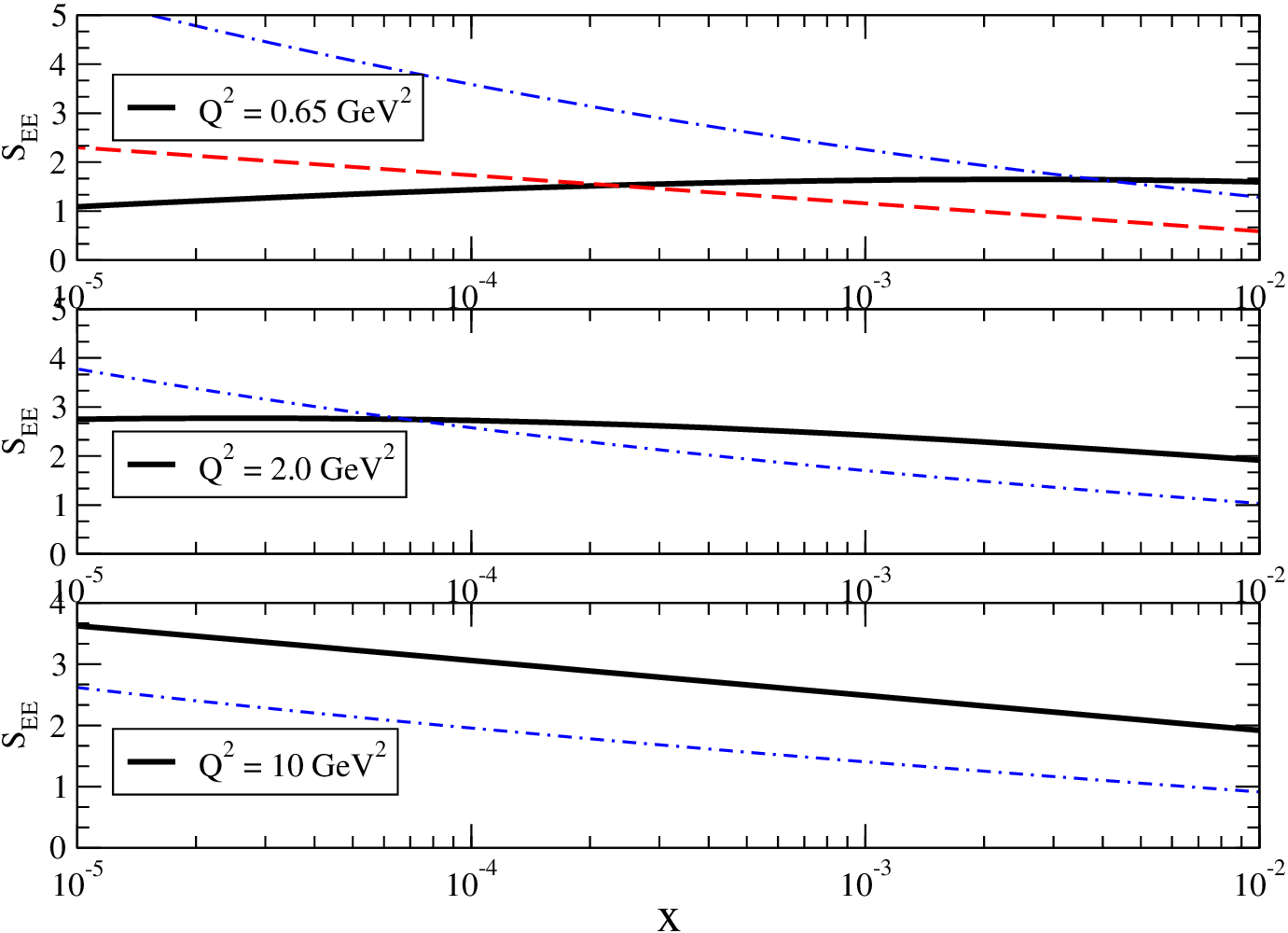} 
	\caption{Entanglement entropy as a function of $x$ for virtualities $Q^2 =0.63,\,2,\,10$ GeV$^2$ in DIS off protons. For the low scale $Q^2=0.63$ GeV$^2$ the maximum entropy at small-$x$ is shown (long-dashed line). The parametric expression $S_{EE}=\ln [r^2Q_s^2(x)]Y$ is also presented (dot-dashed lines).}
	\label{fig:1}
\end{figure}

Here, we will focus on the numerical calculation of the entanglement entropy in the small-$x$ limit both for electron-proton and electron-ion collisions. In Fig. \ref{fig:1} one presents $S_{EE}$ for DIS off proton as a function of $x$ ($10^{-5}\le x\le 10^{-2}$) for representative photon virtualities. We start with a very low scale, $Q^2=0.65$ GeV$^2$, typical of a soft regime which in general can not be addressed by DGLAP evolution starting in an initial hard scale $Q_0^2\sim 2$ GeV$^2$. Notice that the gluon distribution we are using is obtained from the color dipole cross section including parton saturation, which describes successfully the proton structure function, $F_2(x,Q^2)$ at very low-$x$ \cite{Golec-Biernat:2017lfv}. The results for virtualities $Q^2=2$ and $Q^2=10$ GeV$^2$ are also presented. It is very clear the transition from soft to hard scales. Using the parametrization for the saturation scale, $Q_s^2(x)=(x_0/x)^{\lambda}$ (with $\lambda = 0.248$), one verifies that $Q_s^2$ is of order $Q^2=0.63$ GeV$^2$ at $x\lesssim 10^{-3}$. The advantage of using an analytical expression for $xG$ is  to trace back the behavior in terms of scaling variable $\tau = Q^2/Q_s^2$. At $\tau\ll 1$, the series expansion gives $xG\propto Q^4/Q_s^2$ and than $S_{EE} \propto -\log (Q_s^2)$. That is, $S_{EE}\sim \lambda \log(x)$ as viewed at very low $x$. When $\tau =1$, one obtains  $xG\propto [1-(2/e)]Q_s^2$ which leads to $S_{EE}\sim -\lambda \log (x) -1$ and we see in the curve the change in inflection in the transition region $Q^2 \approx Q_s^2$. In the hard regime, where $Q^2\gg Q_s^2$ the asymptotic behavior is given by $xG\propto  Q_s^2(x)$ and $S_{EE}\sim -\lambda \ln(x)$. This is viewed in the plots for $Q^2=2$ GeV$^2$ at larger $x$ and for all $x$ in the case $Q^2=10$ GeV$^2$. 

Here, some comments are in order. The gluon distribution obtained from the unintegrated gluon function shows a valence-like behavior, $x^{\lambda}$ as $x\rightarrow 0$. That is similar to the behavior of the usual DGLAP approach with a valence type parametrization for the gluon PDF at initial scale $Q_0$. However, in last case the pattern fastly disappears with $Q^2$ evolution. The dipole approach includes all twist corrections and then the obtained gluon PDF is somewhat different from the LO DGLAP calculation which is leading twist. In Ref. \cite{Thorne:2005kj}, these features are deeply investigated and a model is proposed for the gluon PDF which at low $Q^2<0.5$ GeV$^2$ behaves as $xG(x,Q^2)\sim Q^2$ and becomes flat in $x$. Same behavior is found also in Kharzeev-Levin-Nardi (KLN) type UGDs \cite{Carvalho:2008ys}. At low $Q^2$ and very small-$x$ it would be interesting to compare our calculation to the analytical expression of $xG_p(x,\mu^2)$ at next-to-leading-order (NLO) level by Jones-Martin-Ryskin-Teubner (JMRT) \cite{Jones:2013pga}. In this case, the parameters of the NLO gluon fit are determined by a global analysis taking into account DESY-HERA data and the LHCb measurements of exclusive $J/\psi$ production in proton-proton collisions (the probed Bjorken-$x$ reaches $x\sim 10^{-6}$, with $\mu^2\simeq m_c^2$, in charmonium photoproduction extracted from ultraperipheral $pp$ collisions).  For sake of comparison, in Fig. \ref{fig:1} we present the result for the $S_{EE}$ using at low $Q^2$ scales the following limit $Q^2\simeq Q_s^2(x)$. This is represented by the long-dashed curve at $Q^2=0.65$ GeV$^2$. In what follows we consider only the kinematical ranges on $Q^2$ where $S_{EE}$ is equal or smaller than its maximum.  For sake of completeness, the parametric expression for the entanglement entropy, Eq (\ref{1p3}) is also presented, using $Q_s^2r^2 = (4Q_s^2/Q^2) + e$  (dot-dashed lines) as discussed before.

The determination of $S_{EE}$ from data was recently done in Ref. \cite{Tu:2019ouv}. For DIS off proton at small-$x$ in DESY-HERA energy range, $\sqrt{s_{ep}}\simeq 225$ GeV, the authors considered Monte Carlo simulations (PYTHIA 6) for the multiplicity distribution in order to obtain the entropy of the final state hadrons, $S_{hadron}$,  and compared it to the entanglement entropy determined by the gluon distribution. The main point is that the $S_{h}$ and the entropy of initial state $S_{EE}$ obey an inequality, $S_h\geq S_{EE}(Y)$, if the second law of thermodynamics applies to entanglement entropy.  For instance, they used the leading order Parton Distribution Function (PDF) set MSTW \cite{Martin:2009iq} and demonstrated that the entropy reconstructed from the final state hadrons is not correlated to $S_{EE}$ at $Q^2=2$ and $Q^2=10$ GeV$^2$. In both virtualities, one has a flat behavior $S_{hadron}\approx 1.5$ for any $\langle x \rangle $ against a powerlike behavior for $S_{EE}$. Our results using a saturated gluon distribution for $Q^2\ge 2$ GeV$^2$ is somewhat similar to those from MSTW PDF presented in Ref. \cite{Tu:2019ouv}, as expected for a kinematic range where $Q^2\geq Q_s^2(x)$. It is argued that DESY-HERA experiment did not cover the kinematic regime where the expression of $S_{EE}$ in terms of gluon distribution applies and the Monte Carlo models do not encode quantum entanglement.  It is expected that the available range for $x$ will be amplied in the proposed  $ep(A)$ colliders like the Large Electron-Hadron Collider (LHeC). For LHeC with energy $\sqrt{s_{ep}}\geq 1$ TeV, DIS kinematics cover $2\times10^{-6}\leq x\leq 0.8$ and $2\leq Q^2\leq 10^5$ GeV$^2$.  

\begin{figure}[t]
		\includegraphics[scale=0.35]{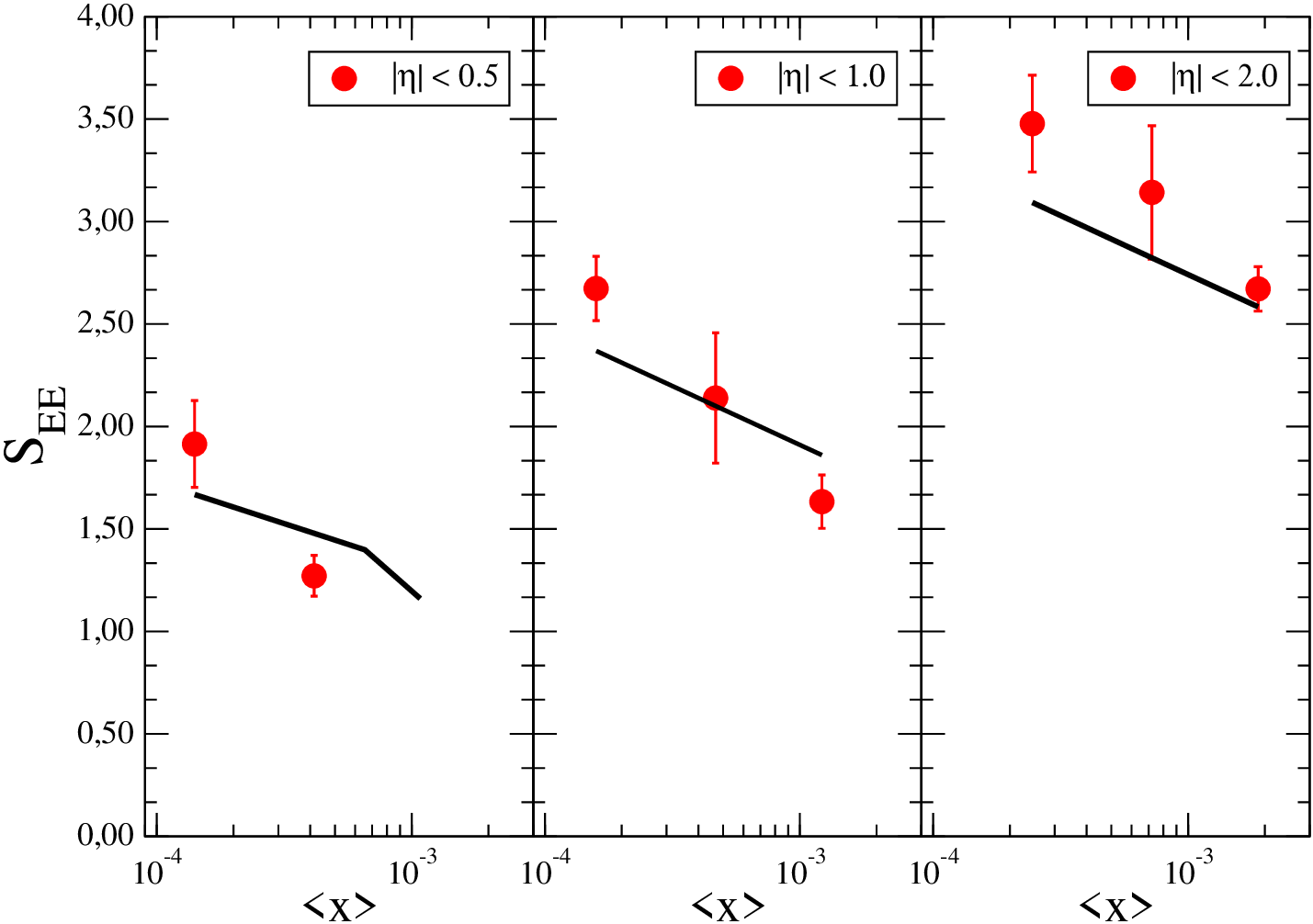} 
	\caption{Entanglement entropy in $pp$ collisions at the LHC, with the final state hadron entropy $S_{had}$ determined in different pseudorapidity ranges (the bins $|\eta |<0.5,1.0, 2.0$ are presented taken from Ref. \cite{Tu:2019ouv}). The numerical result from this  work is represented by the solid lines. }
	\label{fig:2-a}
\end{figure}

Finally, we discuss the case when proton-proton collisions are considered. In Ref. \cite{Tu:2019ouv} the authors modify the multiplicity distribution, $P(N)$, doing an extrapolation in order to reflect a single proton as in $ep$ collisions. The procedure is based on the assumption that final state hadrons are produced coherently by the proton-proton collisions. Moreover, they consider the typical scale in an average $pp$ reaction as being the saturation scale, $Q^2\approx \langle p_{\perp}^2 \rangle\simeq Q_s^2(x)$. Here, we do not argue about the reliability of hypothesis considered in the extraction of $S_{hadron}$ in $pp$ case. In Table \ref{table:1} we present the entanglement entropy given by Eq. (\ref{xgsee}) using the scale $Q^2=Q_s^2(x)$ and following the same procedure proposed in \cite{Tu:2019ouv} to compare it to final state  $S_{hadron}$. A selection on hadron rapidity, $y$, is taken into account based on the different experimental cuts for multiplicity distribution concerning the hadron pseudorapidity, $\eta$. Thus, $S_{hadron}$ is extracted from experimental data from CMS collaboration \cite{Khachatryan:2010nk}, which are consistent with similar measurements done by ATLAS and ALICE collaborations. On the other hand, $S_{EE}=\ln(N_{gluon})$ is obtained computing the number of gluons $N_{gluon}$ by units of rapidity after integration of the gluon PDF over the given rapidity range at a fixed $Q^2$. Specifically, $N_{gluon}=\int_{x_1}^{x_2}[xG(x,Q_s^2)/x]dx$ and $S_{EE}=\ln(N_{gluon})$ is computed for the average $x$, $\langle x\rangle$. In tables I-V of \cite{Tu:2019ouv} are shown the values of the $x$ interval, $[x_1,x_2]$, corresponding to the rapidity range and their average values $\langle x \rangle$. In Table \ref{table:1} we present our results, compared to some extracted values of the final states entropy. 

\begin{table*}[t]
\caption{The entanglement entropy, $S_{EE}$, in proton-proton collisions at the LHC predicted by gluon saturation PDF using procedure from Ref. \cite{Tu:2019ouv}. Some of the extracted values from CMS data are also presented (in parenthesis) \cite{Tu_comm}.}
\begin{tabular}{|c|c|c|c|c|c|}
\hline
 $\sqrt{s_{pp}}$ (TeV) & $|y|< 0.5$ & $|y|< 1.0$ & $|y|< 1.5$ & $|y|< 2.0$ & $|y|< 2.4$ \\
 \hline
 7.00 & 1.668 (1.914 $\pm$ 0.212) &  2.368 (2.673 $\pm$ 0.157) &  2.787 & 3.093 (3.478 $\pm$ 0.236) & 3.291  \\
 2.36 & 1.398 (1.271$\pm$ 0.099)&  2.100 (2.139 $\pm$ 0.318)& 2.517 & 2.823 (3.142 $\pm$ 0.326)& 3.022 \\
 0.90 & 1.160&  1.860 (1.633 $\pm$ 0.130) &  2.277& 2.584 (2.671 $\pm$ 0.108) & 2.784 \\
 \hline
 \end{tabular}
 \label{table:1}
 \end{table*}

 Using $Q^2=Q_s^2$, we obtains an analytical expression for $S_{EE}$, which reads,
 \begin{eqnarray}
 \label{SEESAT}
  S_{EE}(Q^2=Q_s^2) = \ln \left[Q_s^2(x)\right]+ S_0,
 \end{eqnarray}
where $S_0 = \ln[3(e-2)R_p^2/4e\pi \alpha_s ]\simeq 2$ for $\alpha_s=0.2$ and $S_{EE}=S_0$ when $Q_s^2=1$ GeV$^2$. In. Fig. \ref{fig:2-a} we show the entanglement entropy evaluated in this work with the values extracted from the CMS data for the bins $|\eta |<0.5$, $|\eta |<1.0$ and $|\eta |<2.0$. There is a good agreement between the  $S_{EE}$ predicted by the saturation model for the gluon PDF and  the entropy reconstructed from hadron multiplicity at very small-$x$. Interestingly, on the other hand the  usual collinear PDFs give  smaller values for $S_{EE}$ compared to data when the average $\langle x \rangle$ increases. It should be noticed that the larger the $|y|$ interval the bigger the average $\langle x \rangle$, for instance one has $\langle x \rangle = 1.41\cdot 10^{-4}$ for $|y|<0.5$ in contrast to $\langle x \rangle = 3.08\cdot 10^{-4}$ for $|y|<2.4$. The origin of the shortcoming for collinear PDFs can be traced back to the typical powerlike behavior on $x$ even to low scales near the saturation scale $Q^2\sim Q_s^2$. On the other hand, in the saturation limit the saturated gluon distribution considered in this work is basically flat or at most logarithmic.

\subsection{Nuclear entanglement entropy}
\label{sec3b}
\begin{figure}[t]
		\includegraphics[scale=0.35]{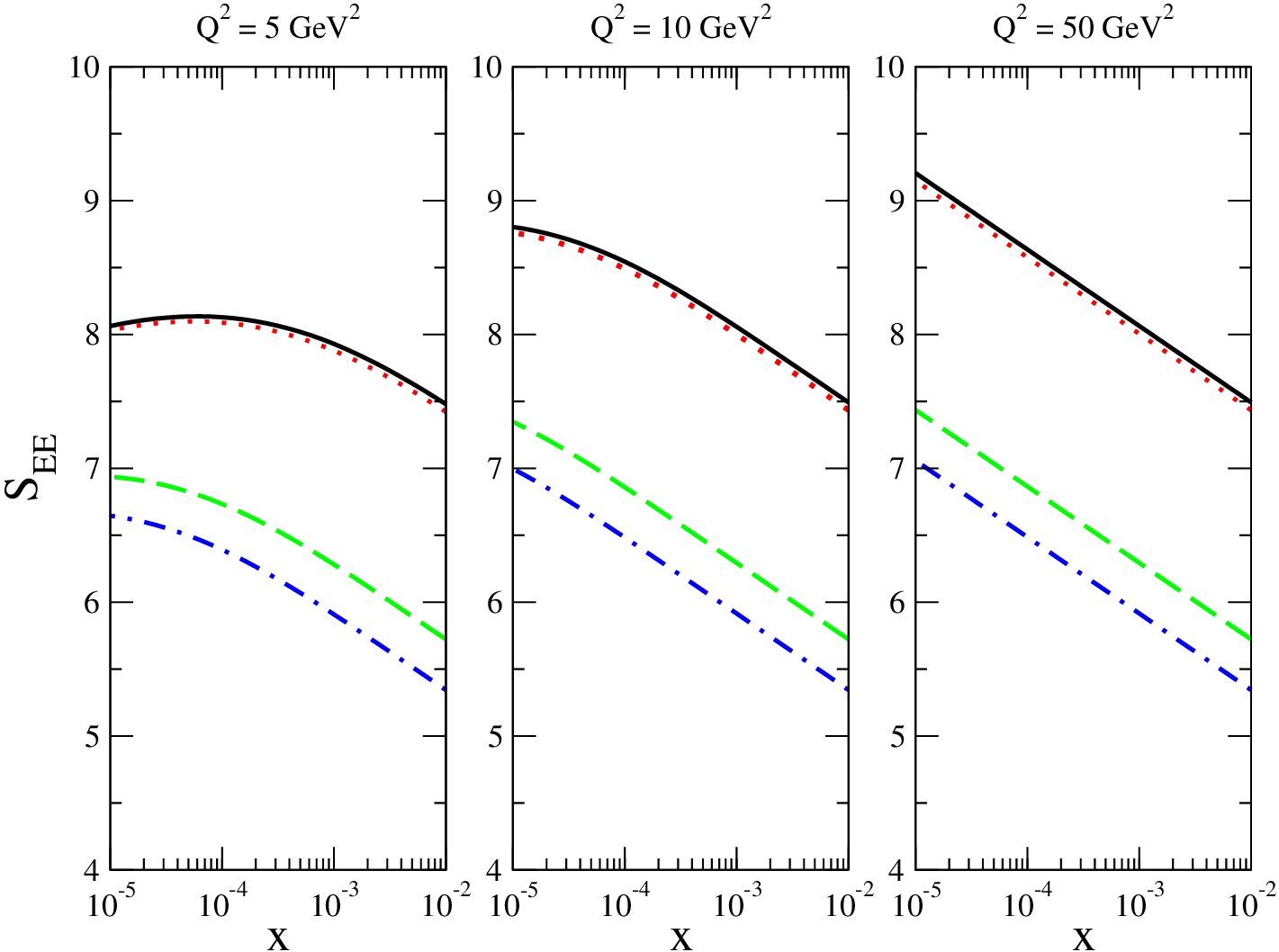} 
	\caption{Nuclear entanglement entropy as a function of $x$ for virtualities $Q^2 =5,\,10,\,50$ GeV$^2$ in DIS off nuclei. For each virtuality, the following nuclides are considered: Pb (solid lines), Au (dotted lines), Ca (long-dashed lines) and Si (dot-dashed lines). }
	\label{fig:2}
\end{figure}

Now we address the entanglement entropy of partons in case of nuclear targets. In order to investigate the entanglement entropy in the case of DIS off nuclei, for simplicity we will consider the geometric scaling property of the parton saturation approaches. That is, the DIS cross section in $eA$ collisions at small-$x$ is directly related to the cross section for  a proton target. The nuclear effects are absorbed in the nuclear saturation scale, $Q_{s,A}^2(x,A)=[A\pi R_p^2/\pi R_A^2]^{\Delta}Q_s^2(x)\sim A^{4/9}Q_s^2(x)$, with $\Delta\simeq 1.27$ \cite{Armesto:2004ud} and the normalization of cross section is rescaled relative to $ep$ by the change $\sigma_A\rightarrow (\pi R_A^2/\pi R_p^2) \sigma_0\sim A^{2/3}\sigma_0$. Here, $R_A\simeq 1.12 A^{1/3}$ fm is the nuclear radius. Therefore, the simplest extension of the gluon distribution in nuclei is given by:
\begin{eqnarray}
 xG_A(x,Q^2) = \frac{3R_A^2}{4\pi\alpha_s}Q_{s,A}^2\left[1-\left(1+\frac{Q^2}{Q_{s,A}^2} \right)e^{-\frac{Q^2}{Q_{s,A}^2}} \right].
\end{eqnarray}

The parametrization based on the color dipole picture and parton saturation formalism is quite reliable and it describes correctly inclusive $\gamma^*p$ and $\gamma^* A$ interactions at small-$x$. In particular, the geometric scaling property described above reproduces without further fitting procedure the experimental data on the energy and centrality dependence of multiplicity of charged particles at RHIC and LHC \cite{Armesto:2004ud}. The main features of the measured ratios of central and semi-central to peripheral $pA$ and $dA$ collisions, $R_{\mathrm{CP}}$, are also roughly described. More recently, the same approach was demonstrated to describe all exclusive processes in $ep$ and $eA$ collisions at small-$x$ like Deeply Virtual Compton Scattering (DVCS) and exclusive vector mesons production. Predictions for exclusive $Z^0$ photoproduction, Timelike DVCS and exclusive dilepton production are presented for instance in Refs. \cite{Ben:2017xny,Machado:2008zv,Machado:2008tp}.

In Fig. \ref{fig:2} we calculate the corresponding nuclear entanglement entropy from the analytical parametrization for the nuclear gluon PDF. We consider the virtualities $Q^2= 5$, 10  and 50 GeV$^2$ and the following nuclei: lead (Pb), gold (Au), calcium (Ca) and silicon (Si). Nuclei Pb and Au are reference for future electron-ion colliders like LHeC and eRHIC. The case $Q^2=2$ is interesting as the nuclear saturation scale (squared) is enhanced by a factor $A^{4/9}$ compared to saturation scale for proton target. This is factor 10 for lead ($A=208$) and 5 for calcium ($A=40$). Therefore, in the model we are using here the scale $Q_{s,A}^2$ is of order 2 GeV$^2$ already at  $x\simeq 10^{-2}$ for Pb and $x\simeq 10^{-3}$ for Ca, whereas in the proton case it occurs at $x\sim 10^{-5}$ (see Fig. \ref{fig:1}).  This means that the $S_{EE}$ will reach to its maximum value for larger value of $x$ compared to DIS off nucleons due to the faster gluon saturation in the nuclear case. We see that  entropy plateau already appears for lead and gold at a sufficiently hard scale $Q^2=5$ GeV$^2$.   

The topic of entanglement entropy and its connection to nuclear shadowing was addressed recently in Ref. \cite{Castorina:2020cro}. The authors  claim that the gluon shadowing is due to a reduction of the entanglement between the observed and unobserved degrees of freedom for gluons in a nucleus compared to those in free nucleon. Specifically, the nuclear entanglement entropy is given by $S_A=A\ln[xG_A(x,Q^2)/A]$, and $S_A/A$ is the entanglement entropy per nucleon ($xG_{N/A}=xG_{A}/A$ is the nuclear gluon density per nucleon). Then, in \cite{Castorina:2020cro} nuclear shadowing is  a direct measure of the variation of the entanglement entropy by nucleon. For two nuclei having atomic number $A$ and $B$, respectively, the nuclear ratio takes the form,
\begin{eqnarray}
 R_g^{A/B}(x,Q^2)=\left(\frac{B}{A}\right)\left[\frac{xG_A(x,Q^2)}{xG_B(x,Q^2)}\right]=\exp \left(\frac{S_A}{A} - \frac{S_B}{B}  \right),
\end{eqnarray}

Accordingly, the number of degrees of freedom per nucleon investigated in DIS in a nucleus of atomic number $A$, $m_A$, is smaller that those for a free nucleon, $m_D$ (gluons in deuterium are considered as those in a free nucleon). In \cite{Castorina:2020cro} one estimates the nuclear entanglement using the Page approach \cite{Page:1993df,Sen:1996ph} for the average entanglement entropy of a subsystem applied to  DIS in a nucleus target.  In such approach, one considers the Hilbert space with dimension $N=mn$ of a quantum bipartite system having dimensions $m$ and $n$, respectively. The Page conjecture provides analytical expressions for the entanglement entropy in both cases $m\leq n$ and $m\geq n$ (see Ref. \cite{Castorina:2020cro} for details). Moreover, it is proposed that antishadowing is connected to the conservation of total entropy, that is $\int_0^1\left[S_A(x)-S_D(x) \right] dx=0$.

\begin{figure}[t]
		\includegraphics[scale=0.35]{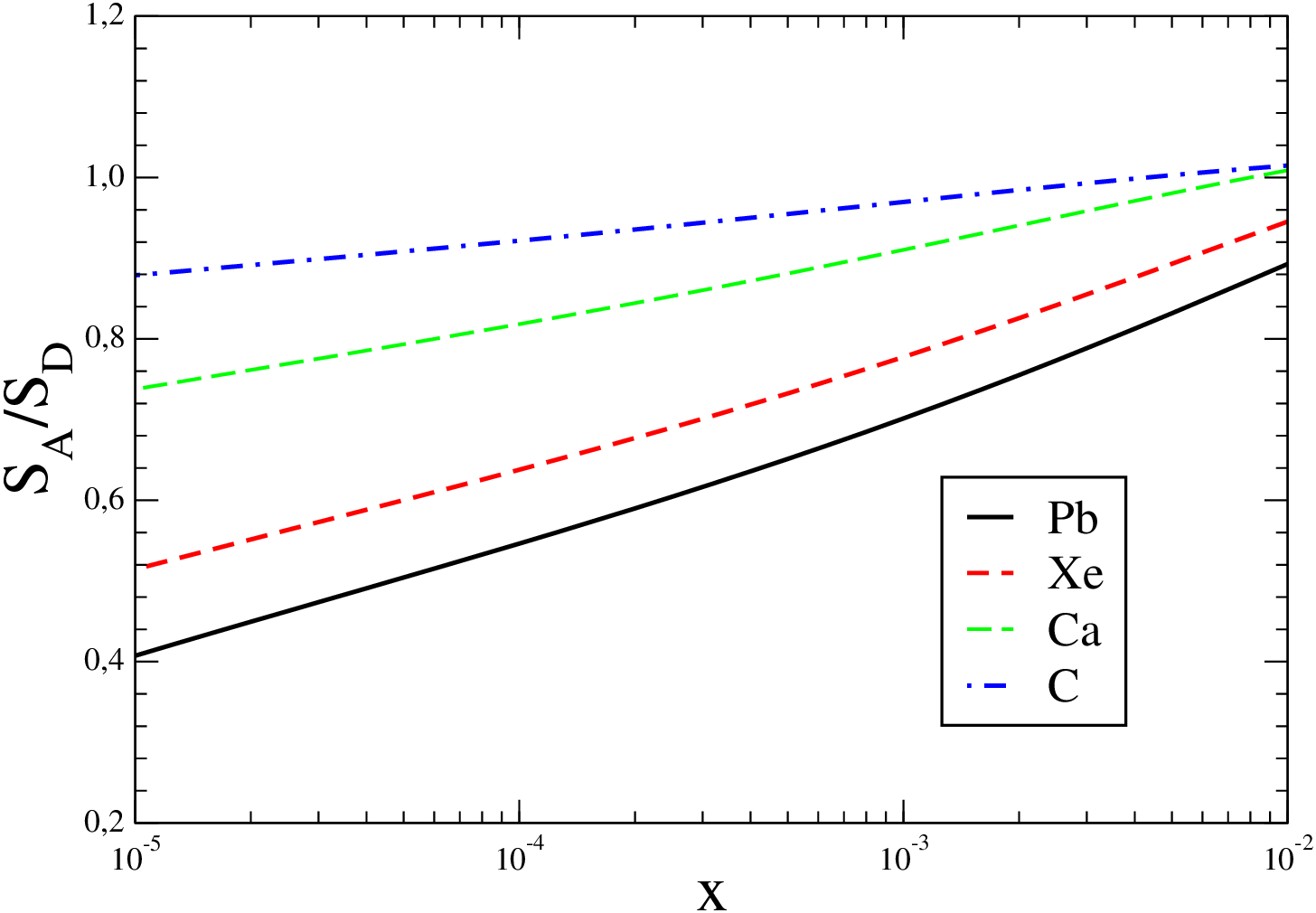} 
	\caption{Ratio $S_A/S_D$ as a function of $x$ at $Q^2=1.7$ GeV$^2$ for different nuclei. It is shown prediction for lead (Pb), xenon (Xe), calcium (Ca) and carbon (C).  }
	\label{fig:3}
\end{figure}

In order to compare our calculations with those in Ref. \cite{Castorina:2020cro}, in Fig. \ref{fig:3} the ratio $S_A/S_D$  is presented as a function of Bjorken-$x$ for a fixed value of virtuality, $Q^2=1.7$ GeV$^2$. We consider the nuclides Pb (solid line), Xe (dashed line), Ca (long dashed line) and C (dot-dashed line). Notice that the ratios obtained in \cite{Castorina:2020cro} are not dependent on $Q^2$ and the degrees on freedom $m_A$ are obtained by fitting the EPPS16 \cite{Eskola:2016oht} output for the gluon shadowing at $Q_0^2\simeq 1.7$ GeV$^2$. Our results for lead and carbon are in agreement to those in  \cite{Castorina:2020cro}, obtaining $S_{Pb}/S_D\simeq 0.5$ and  $S_{C}/S_D\simeq 0.85$ at $x=10^{-4}$ (the same ratios there give 0.3 and 0.7 for equal values of $x$, respectively). The nuclear gluon density we are taking into account describes correctly the nuclear shadowing at small-$x$ for a large variety of nuclei (see Ref. \cite{Betemps:2009da} for the corresponding phenomenology).

The nuclear entropy can be also evaluated in the CGC and Wehrl approaches. Let us take as an example the Wehrl entropy obtained fro  the QCD WW Wigner distribution for gluons.  For sufficiently large $Q^2\gg \tilde{Q}_{s,A}^2$ (leading to a constant $F=-0.248$), where $\tilde{Q}_s$ is the gluon saturation scale in a nucleus,  we will obtain,
\begin{eqnarray}
S_W^A  \approx\frac{C_F}{2\pi \alpha_s}\int_{0}^{\infty}db^2\, 0.248\tilde{Q}_{s,A}^2(x,b),
\end{eqnarray}
where now $\tilde{Q}_{s,A}(x,b)$ is the impact-parameter dependent nuclear gluon saturation scale. There is a rich phenomenology on the determination of nuclear (quark) saturation scale in heavy-ion physics.  Its value can change whether distinct treatments of the nuclear collision geometry are considered. As an example, using a local saturation scale, $Q_{s,A}^2(x,b)=Q_{s,A}^2(x,b=0)T_A(b)$ with $T_A$ being the nuclear thickness function ($Q_{s,p}$ is the saturation scale for protons), and a Gaussian $b$-profile for the proton the relation between $Q_{s,A}$ and $Q_{s,p}$ it was found in Ref. \cite{Salazar:2019ncp}. In the hard sphere approximation for the nuclear density $\rho_A$, one has $Q_{s,A}^2=3A(R_p/R_A)^2Q_{s,p}^2\Theta (b-R_A)$, which gives  $Q_{s,A}^2\approx 2.3 Q_{s,p}^2$ for a lead ($A=208$) nucleus. This means that the nuclear saturation squared is a factor 2 or 3 bigger than for protons and unitarity effects are more pronounced. The expression is quite similar to that employed in our calculation of $xG_A$.  In the hard sphere approximation, the Wehrl entropy for a nucleus is given by,
\begin{eqnarray}
S_W^A  &\approx &\frac{0.248C_F}{2\pi \alpha_s}\int_{0}^{R_A^2}db^2\,\left(\frac{N_c}{C_F}\right)3A\left(\frac{R_p}{R_A}\right)^2Q_{s,p}^2,\nonumber\\
         & = & \frac{0.744N_c S_{\perp}^A}{2\pi^2 \alpha_s}\left(\frac{R_p}{r_0}\right)^2A^{1/3}Q_{s,p}^2(x),
\end{eqnarray}
where $R_A\simeq r_0A^{1/3}$ for large nucleus with $r_0=1.12$ fm. The quantity $S_{\perp}^A =\pi R_A^2$ is the nucleus total transverse area and the nuclear Wehrl entropy behaves as $S_W^A\sim A Q_{s,p}^2(x)=Ae^{\lambda Y}$. The CGC prediction will follow the same trend. Therefore, it can be understood that the nuclear entanglement entropy from CGC formalism and the Wehrl entropy for gluons inside nuclei is additive respect to the hadron ones. This feature is somewhat consistent with the entropy being an extensive variable.  The nuclear entropy proposed in Ref. \cite{Castorina:2020cro} discussed before is also consistent with this picture.
 
\section{Summary}

We have investigated the entanglement entropy in  deep inelastic scattering for $ep$ and $eA$ collisions. The theoretical formalism is based on the von Neumman entropy written in terms of the gluon number as a function of Bjorken-$x$ and photon virtualities $Q^2$. Specifically, we consider an analytical expression for the gluon density in proton related to the parton saturation physics within the color dipole picture. The integrated gluon density, $xG$, is then extracted from the corresponding unintegrated one. The approach is able to describe all the important observables in DIS at small-$x$ and up to intermediate $Q^2\sim 50$ GeV$^2$. Based on geometric scaling property, an extrapolation is done in order to obtain the nuclear gluon density, which also has been tested against nuclear ratios data in $eA$ collisions. The obtained nuclear entanglement entropy is compared to other proposals in literature.  In $ep$ case, it was found that the results are similar to those in Ref. \cite{Tu:2019ouv} with deviations only at very low scales, $Q^2\lesssim 1$ GeV$^2$. The origin of this deviation is traced back to the behavior of gluon PDF below saturation scale, $Q_s(x)$.      In $eA$ case, we analyze the relation between gluon shadowing and the decreasing of the entropy per nucleon proposed in \cite{Castorina:2020cro}. The results corroborate the main results found in that reference. The direct comparison to data is done in Fig. \ref{fig:2-a}, with $S_{EE}$  in agreement with final state hadron entropy  in the rapidity region  extracted from CMS data. The results are similar to those obtained in \cite{Tu:2019ouv} using the usual gluon PDFs like MSTW parametrization not including saturation aspects or higher twist effects. There is some improvement for larger values of average $x$ compared to usual collinear PDFs. This can be understood on the distinct behavior of the proposed saturation model gluon density at the saturation line, $Q^2\approx Q_s^2(s)$.  The main results is that the entanglement entropy at scale $Q^2\approx Q_{s,T}^2$ behaves as $S_{EE}\propto \ln [Q_{s,T}^2(x)])$ for a proton target, $T=p$, as well as a nuclear one, $T=A$.

In summary, our study shed light on the entanglement entropy in hard scattering processes using analytical tools which  could bring a better understanding on the underlying dynamics in a quantum bipartite system. The detailed investigation on the entropy production and the entanglement entropy in these processes are crucial to understand the dynamics of multiparticle production in $pp$ and $AA$ collisions at high energies (see Ref. \cite{Muller:2011ra} for a review). For instance, the thermalization present in those reactions in accelerators like LHC and RHIC could be explained as due to the high degree of entanglement  in the wavefunction of colliding particle \cite{Fries:2008vp,Baker:2017wtt,Feal:2018ptp,Feal:2018zlv}.

\begin{acknowledgments}
We thank Zhoudunming Tu for fruitful discussions and  for providing the calculations of final state hadron entropy presented in Ref. \cite{Tu:2019ouv}.	This work was  partially financed by the Brazilian funding agencies CNPq and CAPES.
\end{acknowledgments}

\end{document}